\documentclass[peerreview]{IEEEtran}

\usepackage{epsf}
\usepackage{amssymb}
\usepackage{graphicx}

\newtheorem{thm}{Theorem}

\newtheorem{definition}[thm]{Definition}
\newtheorem{proposition}[thm]{Proposition}

\newtheorem{algorithm}[thm]{Algorithm}
\newtheorem{example}[thm]{Example}

\begin{document}

\title {The Size of Optimal Sequence Sets for Synchronous CDMA Systems}

\author {Rajesh~Sundaresan,~\IEEEmembership{Senior Member,~IEEE}, and
Arun Padakandla
\thanks{The authors are with the Department of Electrical Communication
Engineering, Indian Institute of Science, Bangalore 560012, India.}
\thanks{This work was supported in part by Qualcomm Inc.,
and the Ministry of Human Resources and Development (MHRD, India)
under Grants Part (2A) Tenth Plan (338/ECE) and (376/ECE).}
\thanks{Material in this paper was presented at the IEEE International
Symposium on Information Theory (ISIT 2004) held in Chicago, USA,
July 2004. } }

\maketitle

\begin{abstract}

The sum capacity on a symbol-synchronous CDMA system having
processing gain $N$ and supporting $K$ power constrained users is
achieved by employing at most $2N-1$ sequences. Analogously, the
minimum received power (energy-per-chip) on the symbol-synchronous
CDMA system supporting $K$ users that demand specified data rates is
attained by employing at most $2N-1$ sequences. If there are $L$
oversized users in the system, at most $2N-L-1$ sequences are
needed. $2N-1$ is the minimum number of sequences needed to
guarantee optimal allocation for single dimensional signaling. $N$
orthogonal sequences are sufficient if a few users (at most $N-1$)
are allowed to signal in multiple dimensions. If there are no
oversized users, these split users need to signal only in two
dimensions each. The above results are shown by proving a converse
to a well-known result of Weyl on the interlacing eigenvalues of the
sum of two Hermitian matrices, one of which is of rank 1. The
converse is analogous to Mirsky's converse to the interlacing
eigenvalues theorem for bordering matrices.

\end{abstract}


\section{Introduction}
\label{Introduction}

Consider a symbol-synchronous code-division multiple access (CDMA)
system. The $k$th user is assigned an $N$-sequence $s_{k} \in
\mathbb{R}^N$ of unit energy, {\it i.e.}, $s_k^t s_k = 1$.  The
processing gain is $N$ chips, and the number of users is $K$, with
$K>N$.  User $k$ modulates the vector $s_k$ by its data symbol $X_k
\in {\mathbb{R}}$ and transmits $X_k s_k$ over $N$ chips. This
transmission interferes with other users' transmissions and is
corrupted by noise. The received signal is modeled by
\[
  Y = \sum_{k=1}^{K} s_k X_k + Z,
\]
where $Z$ is a zero-mean Gaussian
random vector with covariance $I_N$, the $N \times N$ identity
matrix.

We consider two problems already studied in the literature. {\bf
Problem I}: User $k$ has a power contraint $p_k$ units per chip,
{\it i.e.}, $E[X_{k}^{2}] \leq N p_k$. The goal then is to assign
sequences and data rates to users so that the sum of the individual
rates at which the users can transmit data reliably (in an
asymptotic sense) is maximized. The maximum value $C_{sum}$ is
called the sum capacity. Viswanath and Anantharam
\cite{199909TIT_VisAna} show the following. Let $p_{tot} =
\sum_{k=1}^K p_k$.
\begin{itemize}
\item {\em Oversized} users, {\it i.e.}, those capable of transmitting at
large powers relative to other users' power contraints, are best
allocated non-interfering sequences;
\item others are allocated generalized Welch-bound equality (GWBE)
sequences \cite{199909TIT_VisAna};
\item $C_{sum} \leq (1/2) \log (1+p_{tot})$, with equality if and
only if no user is oversized;
\item no user is oversized if $Np_k \leq p_{tot}$ for every user
  $k$.
\end{itemize}

{\bf Problem II}, a dual to Problem I, is one where user $k$ demands
reliable transmission at a minimum rate $r_k$ bits/chip. The goal is
to assign sequences and powers to users so that despite their mutual
interference and noise, each of the users can transmit reliably at
or greater than their required rates, and the sum of the received
powers (energy/chip) at the base-station is minimized. Guess
\cite{200411TIT_Gue} shows results analogous to those of Problem I.
\begin{itemize}
\item {\em Oversized} users, {\it i.e.}, those that demand large rates
  relative to other users' requirements, are best allocated
  non-interfering sequences;
\item others are allocated GWBE sequences;
\item the received sum power is lower bounded by $\exp \left\{ 2 r_{tot} \right\} - 1$, where
$r_{tot} = \sum_{k=1}^K r_k$, with equality if and only if no user
is oversized;
\item no user is oversized if $Nr_k \leq r_{tot}$ for every user
  $k$.
\end{itemize}

Once the optimal sequences are identified and the power or rate
allocated to a user determined, the quantities have to be signaled
to the typically geographically separated users. This
``control-plane'' signaling eats up some bandwidth on the downlink.
This can be a significant fraction of the available resources when
the system is dynamic: users may enter and leave the system, or the
channel may vary as is typical in wireless channels. The sequences
and allocations may need periodic updates. Every such update
requires the transmission to each user of an $N$-vector representing
the sequence and a real number representing the power or rate
allocated. It is therefore of interest to identify schemes that
result in reduced signaling overhead.

In this paper, for both Problems I and II, we come up with a
generalized WBE sequence allocation for the non-oversized users. The
allocation employs at most $2N-L-1$ sequences, where $L$ is the
number of oversized users. To prove this, we only need to restrict
our attention to the case when no user is oversized and show that we
need at most $2N-1$ sequences. The $2N-L-1$ requirement in the
presence of oversized users follows immediately. This upper bound on
the number of sequences enables reduced signaling on the downlink
when there are a large number of users relative to the processing
gain.

Following Guess \cite{200411TIT_Gue}, we show achievability with
such an allocation via a successively canceling decoder. Indeed, if
there are fewer sequences than users, two or more users will be
assigned the same sequence and they will overlap with each other.
Nonlinear receivers, among which the successively canceling decoder
is an example, are therefore necessary. Our algorithm to identify
the sequences and powers or rates allocated requires $O(KN)$
floating point operations and is numerically stable.

The solution to the above problem draws from an interesting result
in matrix theory. If $A$ is a matrix, let $A^*$ denote its Hermitian
conjugate. If $A$ is Hermitian, let $\sigma(A)$ denote its
eigenvalues in decreasing order. The following is a well-known
result due to Weyl (see for e.g., \cite[Section
4.3]{HorJoh_MatrixAnalysis_1999}). If $A$ is an $N \times N$
Hermitian matrix with $\sigma(A) = \lambda = (\lambda_1, \cdots,
\lambda_N)$, and $c \in {\mathbb{C}}^N$, then $\sigma(A + cc^*) =
\left( \hat{\lambda}_1, \cdots, \hat{\lambda}_N \right) =
\hat{\lambda}$ satisfies the interlacing property
\begin{equation}
\label{InterlacingInequality}
\hat{\lambda}_1 \geq \lambda_1 \geq \hat{\lambda}_2 \geq \lambda_2
\geq \cdots \geq \hat{\lambda}_N \geq \lambda_N.
\end{equation}
In this paper, we demonstrate a converse to Weyl's result. We show
that for any Hermitian $A$ with $\sigma(A) = \lambda$, and any real
$\hat{\lambda}$ such that $\lambda$ and $\hat{\lambda}$
satisfy~(\ref{InterlacingInequality}), there exists $c \in
\mathbb{C}^N$ such that $\sigma(A + cc^*) = \hat{\lambda}$. This
converse facilitates a sequence allocation algorithm that, for both
Problems I and II, satisfies the declared properties on the number
of sequences. The above problem falls within the class of structured
inverse eigenvalue problems. Chu \cite{1998xxSIAMReview_Chu}, Chu
and Golub \cite{200201ActaNumer_ChuGol} provide excellent reviews of
similar inverse eigenvalue problems. The problem we have solved
appears to be new.

We then investigate the need for $2N-1$ sequences. While there are
rate requirements (respectively, power constraints) for which fewer
sequences suffice, in general, $2N-1$ is the minimum number required
to meet any set of power (respectively, rate) constraints.

Finally, we show that if some users can be split, $N$ sequences
suffice. This is useful because any set of $N$ orthogonal sequences
will then work. For example, we may employ the standard basis as in
a TDMA system, or a set of Walsh sequences when $N$ is a power of 2
as in a CDMA system. Moreover, this set can be fixed up front. As
the system evolves, it is sufficient to send an index to this set,
thereby making the sequence signaling on the downlink rather easy.
Some users may be signaled more than one index. If there are no
oversized users, at most $N-1$ users are split into exactly two
each, and therefore will need two indices. Because their energy is
now concentrated in a two-dimensional subspace instead of one, the
benefits of spreading (such as robustness to jamming) is obtained to
a lesser degree. The allocation is however optimal; it will either
maximise sum capacity or minimise sum power.

It is perhaps obvious that if multi-dimensional signaling is
allowed, then $N$ orthogonal sequences suffice. Indeed, the goal of
an optimal allocation is to ensure that energy is spread equally in
all dimensions leading to a GWBE allocation. A sufficiently fine
splitting of the users into virtual users with smaller requirements,
coupled with multiple dimensional signaling per user, will achieve
this. The interesting aspects of the above result, however, are an
identification of the number of users that need be split and the
resulting dimensionality of their signaling.

The problem of sequence detection has attracted much attention since
the work of Rupf and Massey \cite{199407TIT_RupMas} who consider the
scenario where users have identical power constraints and the goal
is to maximize the sum capacity. Viswanath and Anantharam
\cite{199909TIT_VisAna} extend the results for users with differing
power constraints. Guess \cite{200411TIT_Gue} studies the dual
problem (Problem II) and obtains sequences that minimize sum power
subject to users' rate constraints. Waldron \cite{200309TIT_Wal}
shows that GWBE sequences are tight frames, a generalization of
bases useful in analyzing wavelet decompositions. Tropp and others
\cite{200411TIT_TroDhiHea} show that sequence design for Problem I
is a structured inverse singular value problem
\cite{200201ActaNumer_ChuGol} and provide a numerically stable
algorithm whose complexity is $O(KN)$ floating point operations. Our
proposed algorithm (first presented in \cite{200406ISIT_Sun}) has
the same $O(KN)$ complexity and numerical stability properties.
Moreover, our algorithm guarantees optimality with at most $2N-1$
sequences and will work for any ordering of users.

Several iterative algorithms for finding optimal sequences have been
proposed. As our focus is on finite-step algorithms, we refer the
interested reader to a work of Tropp and others
\cite{200501TIT_TroDhiHeaStr} and references therein. Another line
of research has been that of sequence design for suboptimal linear
receivers. Viswanath and others in \cite{199909TIT_VisAnaTse} focus
on linear MMSE receivers and identify sequences and powers to meet a
per-user signal to interference ratio (SIR) requirement. Guess
\cite{200304TIT_Gue} extends the results to decision-feedback
receivers. In our work, we place no restriction on the receivers.

The paper is organized as follows. Section \ref{Polymatroid}
provides the preliminaries and states the problems. Section
\ref{ConverseWeylResult} discusses the converse to Weyl's result.
Section \ref{Assignments} provides the new algorithms, shows their
optimality, and verifies that at most $2N-1$ sequences are utilized.
It also discusses the complexity and numerical stability of the
algorithms. Section \ref{numberOfSequences} deals with the need for
$2N-1$ sequences and Section \ref{multiDimensionalSignaling}
discusses the rate-splitting approach to minimize the number of
sequences. Section \ref{conclusions} provides some concluding
remarks.

\section{Preliminaries and Problem Statements}
\label{Polymatroid}

Suppose user $k$ is assigned sequence $s_k$ and is received at power
$p_k$. Let $S$ be the $N \times K$ matrix $[s_1 ~ s_2 ~ \cdots ~
s_K]$. Then, the capacity region \cite{198610Allerton_Ver} can be
written as
\begin{eqnarray}
  \lefteqn{ C(S,p) = \bigcap_{J \subset \{1,\cdots,K\}}
  \left\{ (r_1, \cdots, r_K) \in {\mathbb{R}}_{+}^K : \right.} \nonumber \\
  \label{CapacityFormula1} && ~~~~~~~\left. \sum_{k \in J} r_k \leq \frac{1}{2N} \log \left| I_N
  + \sum_{k \in J} N p_k \cdot s_k s_k^t \right| \right\},
\end{eqnarray}
where $|A|$ denotes the determinant of the matrix $A$, and $r_k$ is
user $k$'s data rate in bits/chip.

The rate vector $r = (r_1, \cdots, r_K)$ is a {\em vertex} of this
capacity region if
\begin{equation}
  \label{IncrementalRate}
  r_k \stackrel{\Delta}{=}
  \frac{1}{2N} \log
  \left| A_k \right|
  -
  \frac{1}{2N} \log
  \left| A_{k-1} \right|, ~ 1 \leq k \leq K,
\end{equation}
where $A_0 \stackrel{\Delta}{=} I_N$, and $A_k \stackrel{\Delta}{=}
\left( A_{k-1} + N p_k \cdot s_k s_k^t \right)$ for $k = 1, \cdots,
K$. The matrix $A_k$ is real, symmetric, and positive-definite.

The vertex (cf.~(\ref{IncrementalRate})) satisfies $r_k \geq 0$ and
$r \in C(S,p)$. These are deduced as follows. Let $\sigma(A_k) =
\left( \lambda_1^{(k)}, \cdots, \lambda_N^{(k)} \right)$. Clearly
$\lambda_1^{(0)} = \cdots = \lambda_N^{(0)} = 1$. Weyl's result
(\ref{InterlacingInequality}) indicates
\[
  \lambda_1^{(k)} \geq \lambda_1^{(k-1)} \geq
  \lambda_2^{(k)} \geq \lambda_2^{(k-1)} \geq \cdots \geq
  \lambda_N^{(k)} \geq \lambda_N^{(k-1)} \geq 1,
\]
which implies that $|A_k| \geq |A_{k-1}|$, and hence $r_k \geq 0$.
Moreover, $r \in C(S,p)$ because this rate point can be achieved via
successive decoding. (Alternatively, $C(S,p)$ is a polymatroid
\cite{199811TIT_TseHan}, and therefore contains all its vertices in
${\mathbb{R}}_{+}^N$).

With the above ideas fixed, let us now re-state Problems I and II.
\begin{itemize}
\item {\bf Problem I} : Given a per user power contraint of $p = (p_1,
  \cdots, p_K)$ where no user is oversized ({\it i.e.}, $N p_k \leq
  p_{tot}$ for every user $k$), find $S$ and $r$ that satisfy $r \in
  C(S,p)$ and $r_{tot} = (1/2) \log (1 + p_{tot})$, the maximum
  sum-rate among all sequence and rate allocations.
\item {\bf Problem II} : Given a per user rate requirement of $r =
  (r_1, \cdots , r_K)$ bits/chip where no user is oversized ({\it i.e.}, $N
  r_k \leq r_{tot}$ for every user $k$), find $S$ and $p$ that
  satisfy $r \in C(S,p)$ and $p_{tot} = \exp \left\{ 2r_{tot} \right\} - 1$, the
  minimum value among all power and sequence allocations.
\end{itemize}

Section \ref{Assignments} identifies the matrices $A_k$ and the
sequence matrix $S$ that solve Problems I and II. Vertices will play
an important role in the solution.

\section{Converse to Weyl's result}
\label{ConverseWeylResult}

\vspace*{.1in}

\begin{proposition} ({\it Converse to Weyl's result})
\label{Interlacing} Let A be an $N \times N$ complex Hermitian
matrix with $\sigma(A) = (\lambda_1, \cdots, \lambda_N)$. Let
$(\hat{\lambda}_1, \cdots, \hat{\lambda}_N)$ be $N$ real numbers
satisfying the interlacing inequality (\ref{InterlacingInequality}).
Then, there exists a $c \in {\mathbb{C}}^N$ such that $\sigma(A +
cc^*) = (\hat{\lambda}_1, \cdots, \hat{\lambda}_N)$. If, in
addition, $A$ is a real, there exists $c \in {\mathbb{R}}^N$ with
the same properties. $\hfill \Box$
\end{proposition}

\vspace*{.1in}

This result is a pleasing dual to an existence result shown by
Mirsky \cite{1958xxJLMS_Mir} for bordered matrices. The problem
falls under the category of additive inverse eigenvalue problems
\cite{1998xxSIAMReview_Chu}, \cite{200201ActaNumer_ChuGol}.
Proposition \ref{Interlacing} however appears to be new. A proof can
be found in Appendix \ref{proofOfInterlacingProposition}.

Proposition \ref{Interlacing} is crucial to this paper. It
guarantees the existence of a program that takes in the matrix $A$,
a set of prescribed eigenvalues $\hat{\lambda}$ that interlace the
eigenvalues of $A$, and outputs $c = c \left( A, \hat{\lambda}
\right)$ such that $A + c c^*$ is the desired perturbation of $A$.
Let us now focus on the nature of $c$ in a special circumstance that
will be of relevance to the sequence design problem.

\vspace*{.1in}

\begin{proposition}
\label{SameSequence} Let A be an $N \times N$ complex Hermitian
matrix with $\sigma(A) = \lambda = (\lambda_1, \cdots, \lambda_N)$.
Let $A$ be unitarily diagonalized by the unitary matrix $U$, {\it
i.e.}, $A = U \Lambda U^*$, where $\Lambda = \mbox{diag} \{ \lambda
\}$. Fix $l$ satisfying $1 \leq l \leq N$. Let $\hat{\lambda}$ be
$N$ real numbers such that $\hat{\lambda}_j = \lambda_j$ for all $j$
except when $j = l$.  Furthermore, let $\hat{\lambda}$ and $\lambda$
satisfy the interlacing property~(\ref{InterlacingInequality}). Then
\begin{enumerate}
\item \label{a1} with $c = \left( \sqrt{\hat{\lambda}_l - \lambda_l} \right)
u_l$, where $u_l$ is the $l$th column of $U$, we have
$\sigma(A+cc^*) = \hat{\lambda}$;
\item \label{a2} $A$ and $(A+cc^*)$ are simultaneously diagonalized by $U$.
\end{enumerate}
$\hfill \Box$
\end{proposition}

\vspace*{.1in}

\begin{proof}
This is easily verified by looking at the assignment in the proof of
Proposition \ref{Interlacing}. Instead of doing this, we give a
direct proof. Let $U = [u_1 ~ \cdots ~ u_N]$, where $u_i \in
\mathbb{C}^N$. Then $A u_l = \lambda_l u_l$, and $u_l^* u_i = 0$ if
$i \neq l$. Hence for every $i \neq l$,
\[
\left( A + c c^* \right) u_i = A u_i + c \left( c^* u_i \right) = A
u_i + 0 = \lambda_i u_i,
\]
and thus $\lambda_i, i \neq l$, are $N-1$ eigenvalues of $A + c
c^*$. Furthermore,
\[
\left( A + c c^* \right) u_l = \lambda_l u_l + \left(
\hat{\lambda}_l - \lambda_l \right) u_l (u_l^* u_l) =
\hat{\lambda}_l u_l,
\]
and thus $\hat{\lambda}_l$ is an eigenvalue of $A + c c^*$. This
proves the first statement.

The eigenvectors for the two Hermitian matrices are the same and
therefore the same unitary matrix (of eigenvectors) diagonalizes
them.
\end{proof}

\section{Sequence Assignments}

\label{Assignments}

Let the rate vector $r$, the sequence matrix $S$, and the power
vector $p$ be such $r \in C(S, p)$. Then

\begin{equation}
\label{ptotConstraint} p_{tot} \geq \exp \left\{ 2 r_{tot} \right\}
- 1,
\end{equation}
where $p_{tot} = \sum_{k=1}^K p_k$, and $r_{tot} = \sum_{k=1}^K
r_k$. Indeed, this can be easily seen through the following sequence
of inequalities. Let $P = \mbox{diag} \{ p \}$, a $K \times K$
diagonal matrix with diagonal entries given by the elements of $p$.
\begin{eqnarray}
p_{tot} & = & \mbox{trace } P \nonumber \\
\label{b1} & = & \mbox{trace } P S^t S \\
\label{b2} & = & \mbox{trace } S P S^t \\
& = & \frac{1}{N} \cdot \mbox{trace } \left( I_N + N S P S^t \right) - 1 \nonumber \\
& = & \frac{1}{N} \cdot \mbox{trace } A_K - 1 \nonumber \\
\label{b3} & \geq  & |A_K|^{1/N} - 1 \\
\label{b4} & \geq & \exp \left\{ 2 r_{tot} \right\} - 1,
\end{eqnarray}
where (\ref{b1}) follows because the diagonal entries of $S^t S$ are
all 1; (\ref{b2}) follows because $\mbox{trace } AB = \mbox{trace }
BA$ when the products $AB$ and $BA$ are well-defined; (\ref{b3})
follows because arithmetic mean of the eigenvalues exceeds the
geometric mean of the eigenvalues; and (\ref{b4}) follows because $r
\in C(S,p)$ which implies that $r_{tot} \leq \frac{1}{2N} \log
\left| A_K \right|$.

For Problem I, given a set of power constraints,
(\ref{ptotConstraint}) indicates that the maximum achievable sum
capacity is upperbounded by $\frac{1}{2} \log \left( 1 + p_{tot}
\right)$, as observed by Viswanath and Anantharam in
\cite{199909TIT_VisAna}. Analogously, for Problem II, given a set of
rate constraints, (\ref{ptotConstraint}) says that the minimum sum
power is lowerbounded by $\exp \left\{ 2 r_{tot} \right\} - 1$. If
these bounds are achieved, then equality in (\ref{b3}) and
(\ref{b4}) indicate that all the eigenvalues of $A_K$ are the same,
and in particular,
\[
A_K = \left(1 + p_{tot} \right) I_N = \exp \{2 r_{tot} \} I_N,
\]
a well-known property of GWBE sequences (\cite[Section
IV]{199909TIT_VisAna}).

\subsection{Algorithm for Problem II}

The signal and interference matrix in the absence of any users is
$A_0$. The matrix with all users taken into account is $A_K$. Let us
now assign sequences in a {\it sequential} fashion, {\it i.e.}, one
user after another.

Loosely speaking, we begin with $A_0$ and its eigenvalues
$\lambda^{(0)} = (1,\cdots,1)$. We then fill up dimensions with
energies from users, one user after another, jumping to the next
dimension when a dimension reaches $\exp \left\{ 2 r_{tot}\right\}$
upon addition of a user. Each time a user is added, say user $k$, we
ensure that $\lambda^{(k-1)}$ and $\lambda^{(k)}$ satisfy the
interlacing property. Indeed, this is guaranteed precisely because
there are no oversized users. Moreover, they differ in at most two
eigenvalues. Upon each addition, we then simply appeal to
Proposition \ref{Interlacing} to get $c_k$ for user $k$. The
sequence $s_k$ is then $s_k = c_k/||c_k||$ and power $p_k = (c_k^t
c_k)/N$.

The following algorithm makes this notion precise. For simplicity,
we assume that users are assigned in the increasing order of their
indices. The order of the users is immaterial as will become
apparent in the proof of correctness of the procedure. Our
assignment will make the given set of rates the {\it vertex} of some
capacity region.

\vspace*{.1in}

\begin{algorithm} Problem II
\label{AlgorithmProblemII}
{\tt
\begin{itemize}
  \item {\bf Initialization}: Set the following:
  \begin{eqnarray*}
    \lambda_n^{(k)} & \leftarrow & 1, ~~ \mbox{ for } 0 \leq k \leq K, ~
    1 \leq n \leq N; \\
    k & \leftarrow & 1; \\
    n & \leftarrow & 1; \\
    \lambda_{\max} & \leftarrow & \exp \left\{ 2r_{tot} \right\}; \\
    A_0 & \leftarrow &I_N.
  \end{eqnarray*}

  \item {\bf Step 1}: ({\it All users are done}). If $k > K$, stop.

  \item {\bf Step 2 - Case (a)}: ({\it Only one eigenvalue changes}). If
  \begin{equation} \label{conditionStep2aP2}
     \lambda_n^{(k-1)} \cdot \exp \left\{ 2 N r_k \right\} <
     \lambda_{\max},
  \end{equation}
  then, set
  \begin{equation} \label{changeEigStep2aP2}
     \lambda_n^{(k)} \leftarrow
     \lambda_n^{(k-1)} \cdot \exp \left\{ 2 N r_k \right\}.
  \end{equation}
  Go to {\bf Step 3}.

  \item {\bf Step 2 - Case (b)}: ({\it Only one eigenvalue changes; dimension $n$ is to be filled
  up}). If
  \begin{equation} \label{conditionStep2bP2}
     \lambda_n^{(k-1)} \cdot \exp \left\{ 2 N r_k \right\} =
     \lambda_{\max},
  \end{equation}
  then, set
  \begin{equation} \label{changeEigStep2bP2}
     \lambda_n^{(j)} \leftarrow \lambda_{\max}, ~~ \mbox{ for } j = k, \cdots,
     K. \\
  \end{equation}
  Now set
  \[
    n \leftarrow n + 1.
  \]
  Go to {\bf Step 3}.

\item {\bf Step 2 - Case (c)}: ({\it Two eigenvalues change; dimension $n$ is to be filled up}).
  If we are in this step, then
  \begin{equation} \label{conditionStep2cP2}
     \lambda_n^{(k-1)} \cdot \exp \left\{ 2 N r_k \right\} >
     \lambda_{\max},
  \end{equation}
  Set the following:
  \begin{eqnarray}
     \label{changeEig1Step2cP2}
     \lambda_n^{(j)} & \leftarrow & \lambda_{\max}, ~~ \mbox{ for } j = k, \cdots,
     K, \\
     \label{changeEig2Step2cP2}
     \lambda_{n+1}^{(k)} & \leftarrow & \frac{\lambda_{n+1}^{(k-1)} \cdot \lambda_n^{(k-1)}
        \cdot \exp \left\{ 2 N r_k \right\} } { \lambda_{\max}}.
  \end{eqnarray}
  Observe that $\lambda_{n+1}^{(k-1)} = 1$. Now set
  \[
    n \leftarrow n + 1.
  \]

\item {\bf Step 3}: ({\it Assign power and sequence for user $k$}).
  Identify the vector
  \[
     c_k = c \left( A_{k-1}, \lambda^{(k)} \right)
  \]
  from Proposition \ref{Interlacing}. Then set
  \begin{eqnarray*}
     s_k & \leftarrow & c_k / ||c_k ||, \\
     p_k & \leftarrow & \left( c_k^t c_k \right) / N.
  \end{eqnarray*}
  A sequence and power are now allocated to user $k$. Finally,
  \begin{eqnarray*}
     A_k & \leftarrow & A_{k-1} + c_k c_k^t, \\
     k & \leftarrow & k+1.
  \end{eqnarray*}
  Go to {\bf Step 1}. $\hfill \Box$

\end{itemize}
}
\end{algorithm}

\vspace*{.1in}

We now make some remarks on the computations and the numerical
stability of the algorithm. The matrices $A_k$ need not be
explicitly computed. It is sufficient to identify and store the
unitary matrices $U_k = \left[ u^{(k)}_1 ~ \cdots ~ u^{(k)}_N
\right]$ that diagonalize $A_k$. Computation of $c \left( A_{k-1},
\lambda^{(k)} \right)$ utilizes only $U_{k-1}$, $\lambda^{(k-1)}$,
and the new $\lambda^{(k)}$, as seen in the proof of Proposition
\ref{Interlacing}. For Step 2($c$) at most two eigenvalues change in
going from $A_{k-1}$ to $A_k$. So the two matrices share $N-2$
eigenvectors and exactly two eigenvectors change. These are computed
via a rotation in the $(n, n+1)$st plane as follows:
\[
  \left[ u^{(k)}_n ~ u^{(k)}_{n+1} \right] =
  \left[ u^{(k-1)}_{n} ~ u^{(k-1)}_{n+1} \right] ~
  \left[
  \begin{array}{cc}
    \alpha & -\beta \\
    \beta  &  \alpha
  \end{array}
  \right],
\]
where
\begin{eqnarray*}
  \alpha & = &
  \sqrt{
  \frac{\left( \hat{\lambda}_n - \lambda_{n+1} \right) \left( \lambda_n - \hat{\lambda}_{n+1} \right)}
  {\left( \hat{\lambda}_n - \hat{\lambda}_{n+1} \right) \left( \lambda_n - \lambda_{n+1} \right) }
  }, \\
  \beta & = & \sqrt{1 - \alpha^2} =
  \sqrt{
  \frac{\left( \hat{\lambda}_{n+1} - \lambda_{n+1} \right) \left( \hat{\lambda}_n - \lambda_n \right)}
  {\left( \hat{\lambda}_n - \hat{\lambda}_{n+1} \right) \left( \lambda_n - \lambda_{n+1} \right) }
  },
\end{eqnarray*}
with $\lambda \stackrel{\Delta}{=} \lambda^{(k-1)}$ and
$\hat{\lambda} \stackrel{\Delta}{=} \lambda^{(k)}$ for simplicity.
Moreover, $p_k = \hat{\lambda}_{n+1} + \hat{\lambda}_n - \left(
\lambda_{n+1} + \lambda_n \right) $, and
\[
c_k = y_n ~ u^{(k-1)}_n + y_{n+1} ~u^{(k-1)}_{n+1},
\]
where
\begin{eqnarray*}
  y_n & = & \sqrt{\frac{\left| \lambda_n - \hat{\lambda}_n \right| \left| \lambda_n - \hat{\lambda}_{n+1} \right|}
                       {\left| \lambda_n - \lambda_{n+1}\right|}}, \\
  y_{n+1} & = & \sqrt{\frac{\left| \lambda_{n+1} - \hat{\lambda}_n \right| \left| \lambda_{n+1} - \hat{\lambda}_{n+1} \right|}
                           {\left| \lambda_n - \lambda_{n+1}\right|}}.
\end{eqnarray*}
These facts can be verified by direct substitution. Furthermore
several assignments in (\ref{changeEig1Step2cP2}) and in the
initialization step can be implicitly assumed. The upshot is that
computation of sequences as well as the new eigenvectors requires
only $O(N)$ floating point operations. Steps 2($a$) and 2($b$) do
not result in a change in eigenvectors. The sequence in these
simpler cases is a scaled version of $u^{(k-1)}_n$ and requires only
$N$ floating point operations for computation. Hence the algorithm's
complexity is $O(KN)$ floating point operations.

The rotation requires identification of $\alpha$ and $\beta$; these
are between 0 and 1 and hence their computation is numerically
stable. Furthermore, it is easy to verify from the interlacing
inequality that
\[
  \max
  \left\{
    \left( \lambda_n - \hat{\lambda}_{n+1} \right),
    \left( \hat{\lambda}_{n+1} - \lambda_{n+1} \right)
  \right\}
  \leq \lambda_n - \lambda_{n+1},
\]
and therefore computations of $y_n$ and $y_{n+1}$ are also
numerically stable. This verifies that the algorithm is numerically
stable.

Before we give the proof of correctness, we illustrate the algorithm
with an example.

\vspace*{.1in}

\begin{example}
Let $N = 2$. Consider four users with rate requirements $r_1, r_2,
r_3$, and $r_4$, such that $r_k \leq r_{tot}/N$, for $k = 1, \cdots,
4$, {\it i.e.}, there are no oversized users.
\end{example}

\vspace*{.1in}

\begin{figure}
\centering
\includegraphics[width=8in]{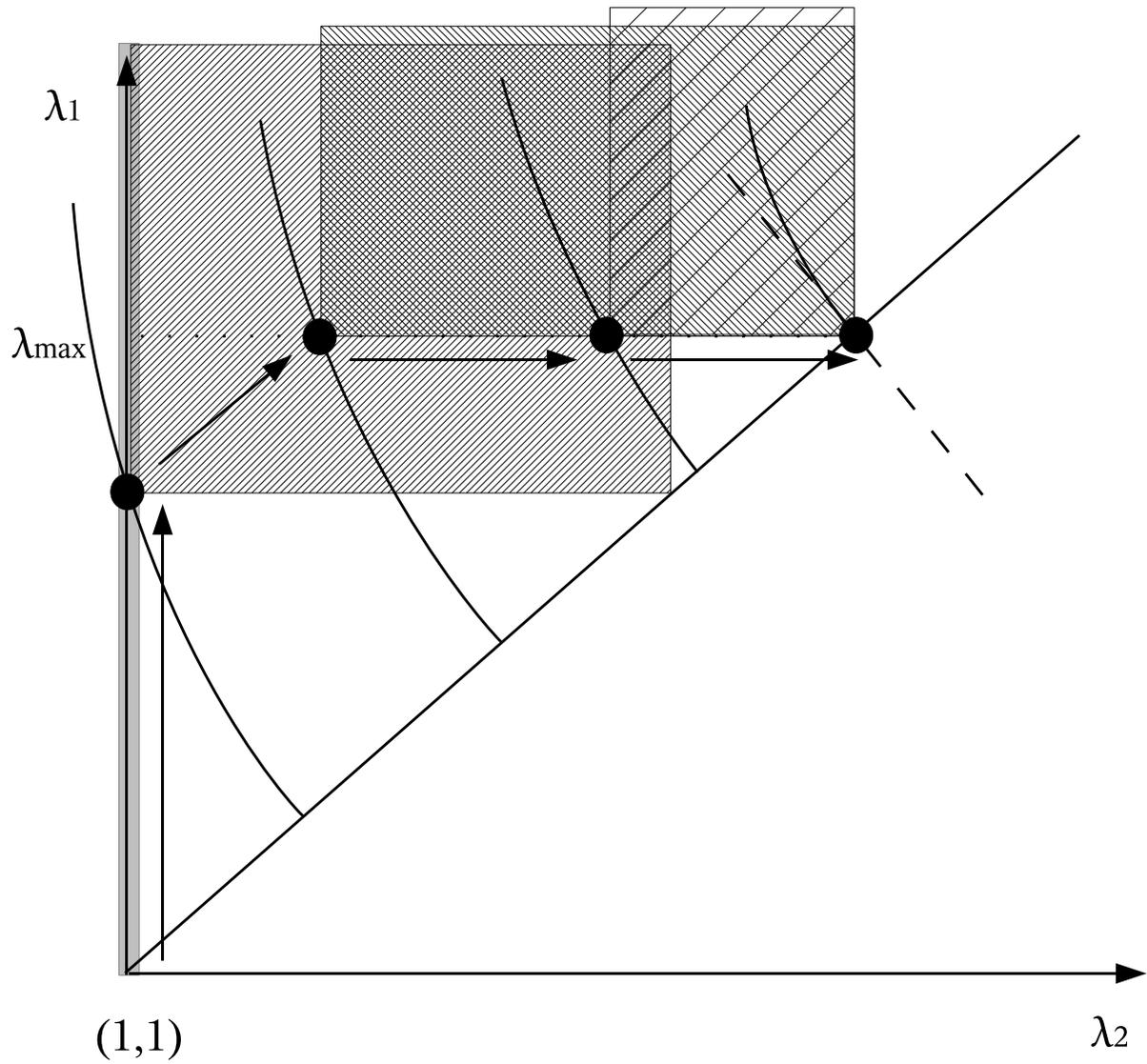}
\caption{Rate constrained sequence allocation algorithm}
\label{fig1}
\end{figure}

Fig. \ref{fig1} represents the eigenvalue space of the matrices
$A_k, ~ k = 0, 1, \cdots, 4$. Our interest is in the upper sector in
the positive quadrant which represents $\lambda_1 \geq \lambda_2
\geq 1$. The algorithm results in a sequence of $\left(
\lambda_1^{(k)}, \lambda_2^{(k)} \right)$, the eigenvalues of $A_k$.

The given rates should be supportable. This places a condition on
the product of the eigenvalues
\[
\lambda_1^{(k)} \lambda_2^{(k)} \geq \exp \left\{ 2N \sum_{j=1}^k
r_j \right\}, ~ k = 1, 2, 3, 4,
\]
{\it i.e.}, after user $k$ is added, the eigenvalue pair should lie
beyond the $k$th hyperbola. Observe that $N + N p_{tot} =
\mbox{trace } A_K = \lambda_1^{(K)} + \lambda_2^{(K)}$ is least, and
therefore $p_{tot}$ is least, if $\lambda_1^{(K)} = \lambda_2^{(K)}
= \lambda_{\max} = \exp \left\{ 2 r_{tot}\right\}$ as is indicated
by the tangent to the outermost hyperbola in the figure. An
allocation that reaches this point is a GWBE allocation. The proof
below shows that this is possible if none of the users are
oversized.

As the users are added sequentially, the interlacing condition says
that the eigenvalues should lie in infinite half-strips shown in
Fig. \ref{fig1}. The condition for $k=1$ is degenerate; it is the
half line represented by $\lambda_2 = 0, ~\lambda_1 \geq 1$.

The first user is added with minimum power to meet the interlacing
and rate conditions. The other users are added to minimize the
number of non-unit eigenvalues (as is the first user). A new
eigenvalue is enlarged only if the previous eigenvalue has reached
$\lambda_{\max}$. The choices of eigenvalue pairs that follow this
rule are indicated in the figure as solid circles. The steps
executed for the four users are $2(a), 2(c), 2(a), 2(b)$,
respectively, for the four users.

Note that the choice is not greedy in the powers that are allocated.
Such a ``power-hungry'' allocation is not guaranteed to reach the
minimum power point, whereas the above algorithm guarantees such an
allocation. $\hfill \Box$

\vspace*{.1in}

\begin{proof} We now give the proof of correctness of Algorithm
\ref{AlgorithmProblemII}. When a new user, say user $k$, is added,
two conditions on the set of eigenvalues of the signal and
interference matrix should be satisfied. One, the added user's rate
should be supportable and hence,
\[
  \frac{1}{2N} \log \left| A_k \right| = \frac{1}{2N} \log
  \prod_{l=1}^N \lambda_l^{(k)} \geq \sum_{j=1}^k r_j;
\]
this imposes a lower bound on the product of the eigenvalues. It is
easy to verify that the algorithm satisfies the above inequality
with equality (see (\ref{vertexAllocation})). Two, the interlacing
eigenvalues condition should hold because the signal and
interference matrix $A_k$ is an as yet unknown rank-1 perturbation
of the matrix $A_{k-1}$.

\vspace*{.1in}

{\em Interlacing eigenvalues} : We first ensure that an execution of
Step 2 results in a set of eigenvalues $\lambda^{(k)} = \left(
\lambda_1^{(k)}, \cdots, \lambda_N^{(k)} \right)$ that interlace the
eigenvalues $\lambda^{(k-1)} = \left( \lambda_1^{(k-1)}, \cdots,
\lambda_N^{(k-1)} \right)$. For a particular $k$, exactly one of the
cases $(a)$, $(b)$, or $(c)$ holds. Consider case $(a)$ or $(b)$.
Then exactly one eigenvalue changes between $\lambda^{(k-1)}$ and
$\lambda^{(k)}$. Indeed, either (\ref{conditionStep2aP2}) or
(\ref{conditionStep2bP2}) holds, and
\begin{eqnarray*}
\lambda^{(k-1)} & = & \left( \underbrace{\lambda_{\max}, \cdots,
\lambda_{\max}}_{n-1}, \lambda_n^{(k-1)}, \underbrace{1, \cdots,
1}_{N-n} \right), \\
\lambda^{(k)} & = & \left( \underbrace{\lambda_{\max}, \cdots,
\lambda_{\max}}_{n-1}, \lambda_n^{(k)}, \underbrace{1, \cdots,
1}_{N-n} \right).
\end{eqnarray*}
To check that the interlacing inequality holds, it is sufficient to
check that
\[
\lambda_{\max} \stackrel{(\alpha)}{\geq} \lambda_n^{(k)}
\stackrel{(\beta)}{>} \lambda_n^{(k-1)} \stackrel{(\gamma)}{\geq} 1.
\]
But this is easily verified. Inequality $(\alpha)$ holds because of
(\ref{conditionStep2aP2}) and (\ref{changeEigStep2aP2}) in case
$(a)$ and because of (\ref{conditionStep2bP2}) and
(\ref{changeEigStep2bP2}) in case $(b)$. Next, inequality $(\beta)$
holds because $r_k > 0$ in both (\ref{changeEigStep2aP2}) and
(\ref{changeEigStep2bP2}). Furthermore, inequality $(\gamma)$ is
preserved at all steps.

To check that the interlacing inequality holds under Step 2 case
$(c)$, note that if this step is executed, (\ref{conditionStep2cP2})
holds, and the eigenvalues are given by
\begin{eqnarray*}
\lambda^{(k-1)} & = & \left( \underbrace{\lambda_{\max}, \cdots,
\lambda_{\max}}_{n-1}, \lambda_n^{(k-1)}, 1, \underbrace{1, \cdots,
1}_{N-n-1} \right), \\
\lambda^{(k)} & = & \left( \underbrace{\lambda_{\max}, \cdots,
\lambda_{\max}}_{n-1}, \lambda_{\max}, \lambda_{n+1}^{(k)},
\underbrace{1, \cdots, 1}_{N-n-1} \right).
\end{eqnarray*}
To see that the interlacing inequality holds, it is sufficient to
check that
\[
\lambda_{\max} \stackrel{(\alpha)}{=} \lambda_n^{(k)}
\stackrel{(\beta)}{\geq} \lambda_n^{(k-1)} \stackrel{(\gamma)}{\geq}
\lambda_{n+1}^{(k)} \stackrel{(\delta)}{>} 1.
\]
This is can be verified as follows. $(\alpha)$ holds because of
$(\ref{changeEig1Step2cP2})$. At all times, our choice of
eigenvalues is such that $1 \leq \lambda_{n}^{(k-1)} \leq
\lambda_{\max}$ and therefore $(\beta)$ holds. To check $(\gamma)$
observe that (\ref{changeEig2Step2cP2}) gives
\[
  \lambda_{n+1}^{(k)} = \frac{\lambda_{n+1}^{(k-1)} \cdot
  \lambda_n^{(k-1)} \cdot \exp \left\{ 2 N r_k \right\} } { \lambda_{\max} } \leq \lambda_n^{(k-1)}
\]
because $\lambda_{n+1}^{(k-1)} = 1$ and $\exp \left\{ 2 N r_k
\right\} \leq \lambda_{\max} = \exp \left\{ 2r_{tot} \right\}$ for
there are no oversized users. Lastly $(\delta)$ follows because of
(\ref{conditionStep2cP2}) and (\ref{changeEig2Step2cP2}).

\vspace*{.1in}

{\em Termination} : After an execution of Step 3, $k$ is the index
of the next user to be allocated, and it is easy to check via
induction and (\ref{changeEigStep2aP2}), or
(\ref{conditionStep2bP2}) and (\ref{changeEigStep2bP2}), or
(\ref{changeEig1Step2cP2}) and (\ref{changeEig2Step2cP2}), that
\begin{equation}
  \label{vertexAllocation}
  \left| A_{k-1} \right| = \exp \left\{ 2 N \sum_{j=1}^{k-1} r_j
  \right\}.
\end{equation}
This ensures that every one of the $K$ users can be added without
the eigenvalues exceeding $\lambda_{\max}$ at any stage and in any
of the dimensions because
\[
  \left| A_{k-1} \right| = \exp \left\{ 2 N \sum_{j=1}^{k-1} r_j
  \right\} \leq \exp \left\{ 2N r_{tot} \right\}
\]
for every $k$. In particular, if at the end of some Step 3 execution
we have $n = N$ for some $k \leq K$, then users $k, k+1, \cdots, K$
can be added without the need for an additional dimension, {\it
i.e.}, Step $2(c)$ will no longer be executed. Moreover, because of
(\ref{vertexAllocation}), we have $\left| A_K \right| = \exp \left\{
2 N r_{tot} \right\}$ and therefore the last Step 2 to be executed
will be Step $2(c)$ thereby correctly filling up the last eigenvalue
at termination.

\vspace*{.1in}

{\em Validity of the sequential procedure} : The previous steps
indicate that the interlacing inequality holds at every step of the
algorithm, and therefore a sequence matrix $S$ and a power
allocation $p$ will be put out by the algorithm. The given $r$, by
design, is a vertex of the capacity region $C(S,p)$. As $C(S,p)$ is
a polymatroid, it contains all its vertices, and therefore $r \in
C(S, p)$. (Alternatively, a successive cancellation receiver
indicates that $r$ is achievable). This implies that the allocations
$S$ and $p$ support the given rate vector $r$, and the proof of
correctness is complete.
\end{proof}

\subsection{Algorithm for Problem I}
We now look at the dual problem of sequence and rate
allocation for sum capacity maximization given users' power
constraints $p = \left( p_1, \cdots, p_K \right)$. Guess
\cite{200411TIT_Gue} explores the duality between these two
problems.

\vspace*{.1in}

\begin{algorithm} Problem I
\label{AlgorithmProblemI}
{\tt
\begin{itemize}
  \item {\bf Initialization}: Set the following:
  \begin{eqnarray*}
    \lambda_n^{(k)} & \leftarrow & 1, ~~ \mbox{ for } 0 \leq k \leq K,
    ~~ 1 \leq n \leq N; \\
    k & \leftarrow & 1; \\
    n & \leftarrow & 1; \\
    \lambda_{\max} & \leftarrow & 1+p_{tot}; \\
    A_0 & \leftarrow & I_N.
  \end{eqnarray*}

  \item {\bf Step 1}: ({\it All users are done}). If $k > K$, stop.

  \item {\bf Step 2 - Case (a)}: ({\it Only one eigenvalue changes}).
  If
  \begin{equation} \label{conditionStep2aP1}
    \lambda_n^{(k-1)}+ N p_k < \lambda_{\max},
  \end{equation}
  then, set
  \begin{equation} \label{changeEigStep2aP1}
    \lambda_n^{(k)} \leftarrow \lambda_n^{(k-1)} + N p_k.
  \end{equation}
  Go to {\bf Step 3}.

  \item {\bf Step 2 - Case (b)}: ({\it Only one eigenvalue changes; dimension $n$ is to be filled up}).
  If
  \begin{equation} \label{conditionStep2bP1}
    \lambda_n^{(k-1)}+ N p_k = \lambda_{\max},
  \end{equation}
  then, set
  \begin{equation} \label{changeEigStep2bP1}
    \lambda_n^{(j)} \leftarrow \lambda_{\max} \mbox{ for } j = k, \cdots, K.
  \end{equation}
  Now set
  \[
    n \leftarrow n + 1.
  \]
  Go to {\bf Step 3}.

  \item {\bf Step 2 - Case (c)}: ({\it Two eigenvalues change; dimension $n$ is to be filled up}).
  If we are in this step, then
  \begin{equation} \label{conditionStep2cP1}
    \lambda_n^{(k-1)}+ N p_k > \lambda_{\max}.
  \end{equation}
  Set the following:
  \begin{eqnarray}
    \label{changeEig1Step2cP1}
    \lambda_n^{(j)} & \leftarrow & \lambda_{\max} \mbox{ for } j = k, \cdots, K;
    \\
    \label{changeEig2Step2cP1}
    \lambda_{n+1}^{(k)} & \leftarrow & \lambda_{n+1}^{(k-1)} + \lambda_n^{(k-1)} +
    Np_k - \lambda_{\max}.
  \end{eqnarray}
  Observe that $\lambda_{n+1}^{(k-1)} = 1$. Set
  \[
    n \leftarrow n + 1.
  \]

  \item {\bf Step 3}: ({\it Assign rate and sequence for user $k$}).
  Identify the vector
  \[
    c_k = c \left( A_{k-1}, \lambda^{(k)} \right)
  \]
  from Proposition~\ref{Interlacing}. Then set
  \begin{eqnarray*}
     A_k & \leftarrow & A_{k-1} + c_k c_k^t, \\
     s_k & \leftarrow & c_k / \| c_k \|, \\
     r_k & \leftarrow & \frac{1}{2N} \log \left| A_k \right| - \frac{1}{2N} \log \left|
     A_{k-1} \right|.
  \end{eqnarray*}
  A sequence and rate are now allocated to user $k$. Finally,
  \[
    k \leftarrow k+1.
  \]
  Go to {\bf Step 1}. $\hfill \Box$
\end{itemize}
}
\end{algorithm}

\vspace*{.1in}

We now illustrate the algorithm with an example.

\vspace*{.1in}

\begin{example}
\label{exproblem1} Let $N = 2$. Consider a numerical example with
four users having power limits $p = (p_1, p_2, p_3, p_4) = (2, 2, 3,
1)$.
\end{example}

\vspace*{.1in}

\begin{figure}
\centering
\includegraphics[width=8in]{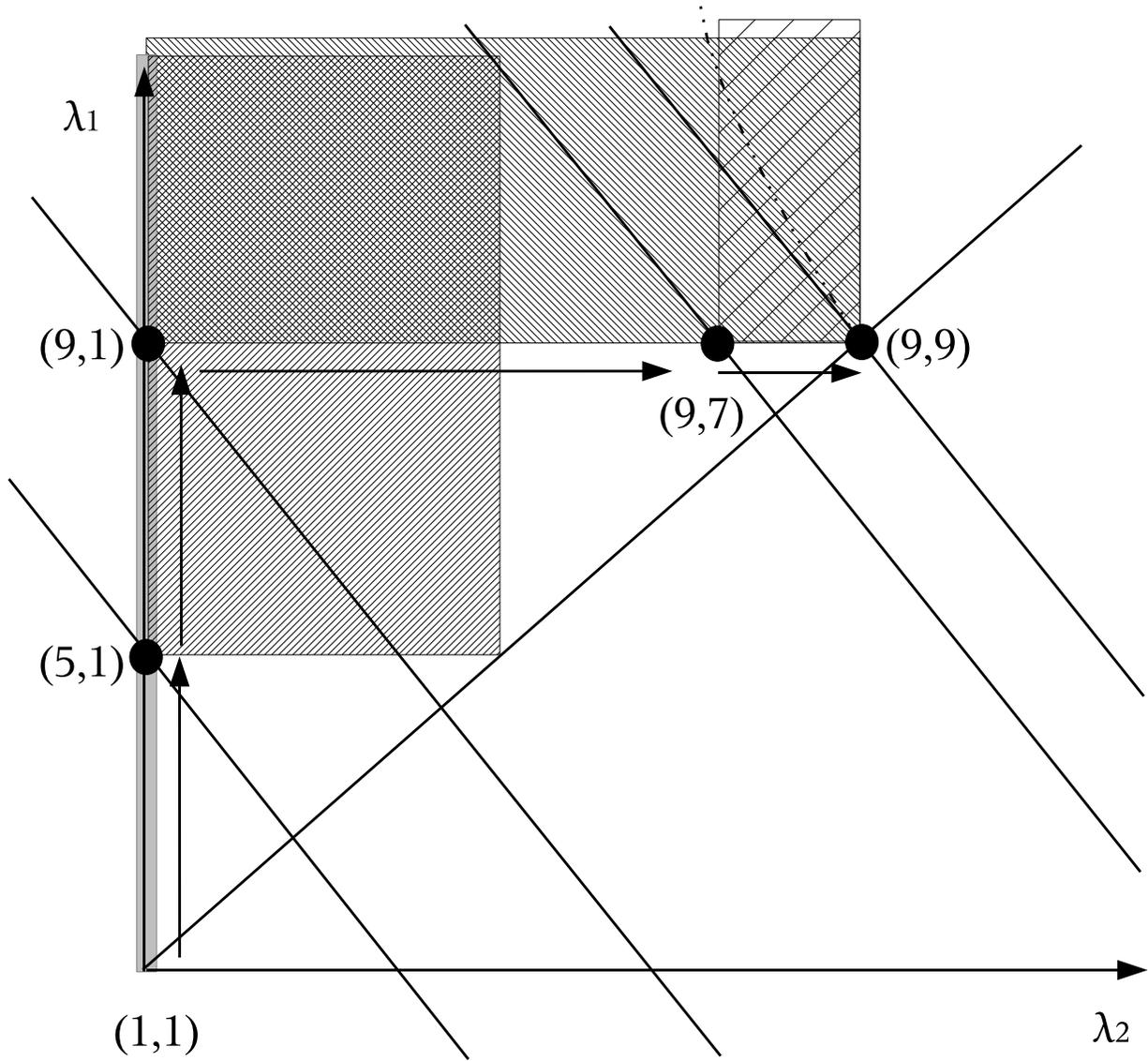}
\caption{Power constrained sequence allocation algorithm}
\label{fig2}
\end{figure}

It is easy to check that no user is oversized. Refer to Fig.
\ref{fig2}, the eigenvalue space of the matrices $A_k$ for $k = 0,
1, \cdots, 4$. Because $\mbox{trace } A_k = N + N \sum_{j=1}^k p_j$,
the eigenvalues are constrained to lie within the confines of the
four triangular regions bounded by
\[
  \lambda_1 \geq 1, ~~ \lambda_2 \geq 1, ~~ \lambda_1 + \lambda_2 \leq
  N + N \sum_{j=1}^{k} p_j
\]
as given in the figure. The last bounding line $\lambda_1 +
\lambda_2 = N + N p_{tot}$ is a tangent to the region
\[
  \lambda_1 \lambda_2 \geq \left( 1 + p_{tot} \right)^N
\]
at the point $\lambda_1 = \lambda_2 = 1 + p_{tot}$. If this point
can be reached, then we have a GWBE allocation and the sum rate is
\[
  \frac{1}{2} \log (1 + p_{tot}) = \frac{1}{2} \log 9,
\]
the maximum among all allocations meeting the power constraints
given by the aforementioned tangent line.

The algorithm ensures that at each step the eigenvalues interlace
and the power constraints met. As before the number of non-unit
eigenvalues are kept at a minimum at each step. The sequence of
eigenvalue pairs are
\begin{eqnarray*}
  (\lambda_1^{(1)}, \lambda_2^{(1)}) & = & (5, 1), \\
  (\lambda_1^{(2)}, \lambda_2^{(2)}) & = & (9, 1), \\
  (\lambda_1^{(3)}, \lambda_2^{(3)}) & = & (9, 7), \mbox{ and}  \\
  (\lambda_1^{(4)}, \lambda_2^{(4)}) & = & (9, 9)
\end{eqnarray*}

The interesting aspect of this numerical example is that users 1 and
2 share a sequence, say $s_1$, and users 3 and 4 share another
sequence, say $s_3$, and moreover, $s_1^t s_3 = 0$. This completely
separates users 1 and 2 from users 3 and 4, attains the minimum
$\frac{1}{2} \log \left( 1 + p_{tot} \right)$ bound, and yet gives
the benefits of spreading. Any pair of orthogonal vectors suffice.
We explore this possibility further in Section
\ref{multiDimensionalSignaling} and show how to obtain orthogonal
sequences under mild rate splitting. $\hfill \Box$

\vspace*{.1in}

\begin{proof} We now give the proof of correctness of Algorithm
\ref{AlgorithmProblemI}. The proof parallels the proof of
correctness of Algorithm \ref{AlgorithmProblemII}. We provide it
here only to highlight the steps that rely on the absence of
oversized users.

When a new user, say user $k$, is added, the power constraint
\[
  \mbox{trace } A_k = \sum_{l=1}^N \lambda_{l}^{(k)} \leq \sum_{j=1}^k
  p_k,
\]
and the interlacing eigenvalue condition are met. The power
constraint is in fact met with equality. The rate assigned to this
user is given by (\ref{IncrementalRate}). We now verify these
statements.

\vspace*{.1in}

{\em Interlacing eigenvalues} : An execution of Step 2 results in
$\lambda^{(k)}$ that interlace $\lambda^{(k-1)}$. Indeed, for a
particular $k$, exactly one of cases $(a)$, $(b)$, or $(c)$ is
executed. Consider case $(a)$ or $(b)$. Then exactly one eigenvalue
changes between $\lambda^{(k-1)}$ and $\lambda^{(k)}$, and
\begin{eqnarray*}
\lambda^{(k-1)} & = & \left( \underbrace{\lambda_{\max}, \cdots,
\lambda_{\max}}_{n-1}, \lambda_n^{(k-1)}, \underbrace{1, \cdots,
1}_{N-n} \right), \\
\lambda^{(k)} & = & \left( \underbrace{\lambda_{\max}, \cdots,
\lambda_{\max}}_{n-1}, \lambda_n^{(k)}, \underbrace{1, \cdots,
1}_{N-n} \right).
\end{eqnarray*}
To check that the interlacing inequality holds, it is sufficient to
check that
\[
\lambda_{\max} \stackrel{(\alpha)}{\geq} \lambda_n^{(k)}
\stackrel{(\beta)}{>} \lambda_n^{(k-1)} \stackrel{(\gamma)}{\geq} 1.
\]
But this is easily verified. Inequality $(\alpha)$ holds because of
(\ref{conditionStep2aP1}) and (\ref{changeEigStep2aP1}), or
(\ref{conditionStep2bP1}) and (\ref{changeEigStep2bP1}), as the case
may be. Inequality $(\beta)$ holds because in
(\ref{changeEigStep2aP1}) or in (\ref{conditionStep2bP1}) we have
$p_k > 0$. The algorithm is such that $(\gamma)$ is preserved at all
steps.

We now verify the interlacing inequality holds under Step $2(c)$. As
a consequence of (\ref{conditionStep2cP1}), we have
\begin{eqnarray*}
\lambda^{(k-1)} & = & \left( \underbrace{\lambda_{\max}, \cdots,
\lambda_{\max}}_{n-1}, \lambda_n^{(k-1)}, 1, \underbrace{1, \cdots,
1}_{N-n-1} \right), \\
\lambda^{(k)} & = & \left( \underbrace{\lambda_{\max}, \cdots,
\lambda_{\max}}_{n-1}, \lambda_{\max}, \lambda_{n+1}^{(k)},
\underbrace{1, \cdots, 1}_{N-n-1} \right).
\end{eqnarray*}
To see that the interlacing inequality holds, it is sufficient to
check that
\[
\lambda_{\max} \stackrel{(\alpha)}{=} \lambda_{n}^{(k)}
\stackrel{\beta}{\geq} \lambda_n^{(k-1)} \stackrel{(\gamma)}{\geq}
\lambda_{n+1}^{(k)} \stackrel{(\delta)}{>} 1.
\]
This can be verified as follows. Equality $(\alpha)$ holds because
of the assignment in (\ref{changeEig1Step2cP1}) for $j = k$. At all
times, our choice of eigenvalues is such that $1 \leq
\lambda_{n}^{(k-1)} \leq \lambda_{\max}$ and therefore $(\beta)$
holds. To check $(\gamma)$ observe that (\ref{changeEig2Step2cP1})
gives
\[
  \lambda_{n+1}^{(k)} = \lambda_{n+1}^{(k-1)} + \lambda_n^{(k-1)} +
    Np_k - \lambda_{\max} \leq \lambda_n^{(k-1)}
\]
because $\lambda_{n+1}^{(k-1)} = 1$ and $ 1 + N p_k \leq
\lambda_{\max} = 1 + p_{tot}$ for there are no oversized users.
Lastly $(\delta)$ follows because of (\ref{conditionStep2cP1}), the
assignment in (\ref{changeEig2Step2cP1}), and the equality
$\lambda_{n+1}^{(k-1)} = 1$.

\vspace*{.1in}

{\em Termination} : After an execution of Step 3, user $k$ is the
next user to be allocated. As before, it is easy to check via
induction, and (\ref{changeEigStep2aP1}), or
(\ref{conditionStep2bP1}) and (\ref{changeEigStep2bP1}), or
(\ref{changeEig1Step2cP1}) and (\ref{changeEig2Step2cP1}), as the
case may be, that
\[
  \mbox{trace } A_{k-1} = N + N \sum_{j=1}^{k-1} p_j,
\]
{\it i.e.}, the power constraints of the just added user is
respected. This also ensures that every one of the $K$ users can be
added because
\[
  \mbox{trace } A_{k-1} = N + N \sum_{j=1}^{k-1} p_j \leq N + N p_{tot}
  = N \lambda_{\max}
\]
for every $k$. In particular, if at the end of some Step 3 we have
$n = N$ for some $k \leq K$, then users $k, k+1, \cdots, K$ can be
added without the need for an additional dimension, {\it i.e.}, Step
$2(c)$ will no longer be executed.

\vspace*{.1in}

{\em Validity of the sequential procedure} : The previous steps
indicate that the interlacing inequality holds at every step of the
algorithm, and therefore a sequence matrix $S$ and a rate allocation
$r$ will be put out by the algorithm. The obtained $r$, by design,
is a vertex of the capacity region $C(S,p)$. $C(S,p)$ is a
polymatroid; it contains all its vertices, and therefore $r \in C(S,
p)$.
\end{proof}

\section{Number of sequences}

\label{numberOfSequences}

In this section, we study the number of sequences needed to design
systems that solve Problems I and II.

\subsection{$2N - 1$ sequences suffice}

We first show that $2N - 1$ sequences suffice regardless of the
number of users $K$. If $K > 2N-1$, then there are fewer sequences
assigned than users and therefore some users share the same
sequence; they will completely overlap with one other. But a
successive cancelation receiver enables us to receive data from all
such users if powers or rates are suitably assigned.

\vspace*{.1in}

\begin{thm}
There is an optimal sequence allocation with at most $2N -1$
distinct sequences for both Problems I and II. Furthermore, when
there are $L$ oversized users, there is an optimal sequence
allocation with at most $2N - L - 1$ distinct sequences. $\hfill
\Box$
\end{thm}

\vspace*{.1in}

\begin{proof}
Suppose that there are no oversized users. We will show that
Algorithms \ref{AlgorithmProblemII} and \ref{AlgorithmProblemI}
result in a design with the desired property. In both algorithms, a
new sequence is put out if either

\begin{itemize}

\item Step $2(c)$ is executed, {\it i.e.}, a ``break-out'' in a new
dimension $n+1$ occurs, or

\item if one of Step $2(a)$ or Step $2(b)$ is executed for the first
time in a new dimension $n$, {\it i.e.}, $n$ was updated when the
previous user was added.

\end{itemize}

The former of these conditions happens at most $N-1$ times because
it results in a break-out in a new dimension and there are at most
$N$ dimensions. The latter happens at most $N$ times because in any
dimension there is only one first step. All subsequent steps in the
same dimension do not result in a new sequence because of
Proposition \ref{SameSequence}. This ensures that the algorithm puts
out a sequence matrix with at most $2N - 1$ sequences.

Suppose now that there are $L$ oversized users. These users should
be assigned $L$ sequences, orthogonal to each other, and orthogonal
to the space spanned by the sequences assigned to the remaining
users. There are $N-L$ dimensions available for the remaining $K-L$
nonoversized users. An appeal to the case with no oversized users
results in at most $2(N-L) - 1$ for these $K-L$ users. In total,
therefore, we need at most $L + 2(N-L) -1 = 2N - L - 1$ sequences.
\end{proof}

\vspace*{.1in}

\subsection{$2N - 1$ sequences are necessary for one-dimensional signaling}

We now provide a converse to the results in the previous subsection.

\vspace*{.1in}

\begin{thm}
For Problem I, given $N$, there exist $K > N$ and power constraints
$p = (p_1, \cdots, p_K)$ with no user oversized, such that any rate
and sequence allocation with $2N - 2$ or fewer sequences results in
a sum capacity
\[
  r_{tot} < \frac{1}{2} \log \left( 1 + p_{tot} \right).
\]

Similarly, for Problem II, given $N$, there exist $K > N$ and rate
requirements $r = (r_1, \cdots, r_k)$ with no user oversized, such
that any power and sequence allocation with $2N - 2$ or fewer
sequences results in a minimum power
\[
  p_{tot} > \exp \left\{ 2 r_{tot} \right\} - 1.
\]
$\hfill \Box$
\end{thm}

\vspace*{.1in}

\begin{proof}
Consider Problem I. Consider $K = 2N - 1$ users with a symmetric
power constraint
\[
  p = \left( \underbrace{\frac{p_{tot}}{K}, \frac{p_{tot}}{K}, \cdots, \frac{p_{tot}}{K}}_{2N - 1} \right).
\]
Let $\left[ s_1, s_2, \cdots, s_K \right]$ be any sequence
assignment with $2N -2$ or fewer sequences, and $(r_1, \cdots, r_K)$
a rate assignment. At least two users share a sequence. Without loss
of generality we may assume that these are users 1 and 2, {\it
i.e.}, $s_1 = s_2$. The basic idea is to show that users 1 and 2 put
together make an oversized {\it compound} user with power constraint
$p_1 + p_2$ in a $2N - 2$ user system. Indeed,
\[
  p_1 + p_2 = \frac{2 p_{tot}}{K } = \frac{2 p_{tot}}{2 N - 1} >
  \frac{p_{tot}}{N},
\]
which makes the compound user oversized. The sum capacity of the
system is therefore strictly smaller than $\frac{1}{2} \log \left( 1
+ p_{tot} \right)$. We now formalize this idea.

$A_2$ is given by
\[
  A_2 = I_N + N p_1 s_1 s_1^t + N p_2 s_2 s_2^t = I_N + \frac{2 N p_{tot}}{K} s_1
  s_1^t,
\]
with eigenvalues
\[
  \lambda^{(2)} = \left( 1+ \frac{2 N p_{tot}}{K}, \underbrace{1, \cdots,
   1}_{N-1} \right),
\]
and therefore
\[
  \lambda_1^{(K)} \geq \lambda_1^{(2)} = 1+\frac{2 N p_{tot}}{K} = 1+\frac{2 N p_{tot}}{2N - 1}
  > 1+p_{tot},
\]
where the first inequality is a consequence of Weyl's result applied
to $A_K = A_2 + \sum_{k = 3}^K N p_k s_k s_k^t$.

Consequently, we have the following sequence of inequalities:
\begin{eqnarray}
   \label{powerTraceRelation}
   \frac{1}{2} \log \left( 1 + p_{tot} \right) & = & \frac{1}{2} \log
   \left( \frac{ \mbox{trace } A_K }{N} \right) \\
   \label{AMGMInequality}
   & > & \frac{1}{2N} \log \left| A_K \right| \\
   \label{rateCondition}
   & \geq & r_{tot},
\end{eqnarray}
where (\ref{powerTraceRelation}) holds because the power constraint
implies $\mbox{trace } A_K = N + N p_{tot}$, (\ref{AMGMInequality})
holds because the arithmetic mean of the eigenvalues is {\em
strictly} larger than their geometric mean, a consequence of the
fact that one of the eigenvalues $\lambda_1^{(K)}$ is strictly
larger than the arithmetic mean of the eigenvalues, and
(\ref{rateCondition}) is a necessary condition to support the rate
vector $r$. This proves the first part of the theorem.

The proof for Problem II is very similar. Once again, we look at $K
= 2N -1$ users with symmetric rate requirements $r = \left(
\frac{r_{tot}}{K}, \cdots, \frac{r_{tot}}{K} \right)$, and make
users 1 and 2 share a sequence. Then users 1 and 2 put together form
an oversized compound user because
\[
r_1 + r_2 = \frac{2 r_{tot}}{K} = \frac{2 r_{tot}}{2 N - 1} >
\frac{r_{tot}}{N}.
\]
Now observe that
\[
  \sigma(A_2) = \left( \lambda_1^{(2)}, \underbrace{1, \cdots, 1}_{N-1} \right)
\]
and therefore to support the rate of the two users we need
\[
  \frac{1}{2N} \log  \lambda_1^{(2)} = \frac{1}{2N} \log
  \left| A_2 \right| \geq r_1 + r_2
  > \frac{r_{tot}}{N}
\]
which leads to
\[
  \lambda_1^{(K)} \geq \lambda_1^{(2)} > \exp \left\{ 2 r_{tot}
  \right\}.
\]
The same sequence of inequalities, (\ref{powerTraceRelation}),
(\ref{AMGMInequality}), and (\ref{rateCondition}), once again hold.
The strict inequality in (\ref{AMGMInequality}) holds because the
largest eigenvalue is now strictly bigger than the geometric mean of
the eigenvalues. Thus $p_{tot} > \exp \left\{2 r_{tot} \right\} -
1$, which completes the proof of the second part of the theorem.
\end{proof}

\section{Two Dimensional signaling and sufficiency of $N$ orthogonal sequences}
\label{multiDimensionalSignaling}

Our goal in this section is to show how to achieve sum capacity
(respectively minimum sum power) by using at most $N$ orthogonal
sequences. Thus far, each user has been confined to signal along a
single dimension. In some cases, as in Example \ref{exproblem1}, it
is possible to get an optimal assignment with $N$ orthogonal
sequences. In general we need $2N-1$ sequences to achieve sum
capacity or minimum power. However, by letting a few users' signals
span two dimensions instead of one, it is possible to achieve
optimality with $N$ orthogonal sequences. We start with the
following definition.

\vspace*{.1in}

\begin{definition}
A vector $x=(x_1,x_2,\cdots,x_K)$ has a {\it symmetric sum partition
of size $N$}, if there is a partition of the users
$\left\{1,2,\cdots,K\right\}$ into $N$ subsets $S_1,S_2,\cdots,S_N$,
such that
\begin{equation}
 \label{sspdefn} \sum_{k \in S_n} x_k ~ = ~ \frac{1}{N}\sum_{k = 1}^{K}x_k
 ~ = ~
 \frac{x_{tot}}{N},
\end{equation}
for $n{=}1,2,\cdots,N$. The subsets  $S_1,S_2,\cdots,S_N$ will be
referred to as the {\it symmetric sum partition}.
\end{definition}

\vspace*{.1in}

{\em Remark} : If $x$ has a symmetric sum partition of size $N$, no
user is oversized. This is because, any user $k'$ belongs to $S_n$,
for some $n$, and (\ref{sspdefn}) implies $x_{k'}$ $\leq  \sum_{k
\in S_n}x_k {=} {x_{tot}}/{N}$. Note that the power constraint
vector $p$ of Example \ref{exproblem1} has a symmetric sum partition
of size 2.

\vspace*{.1in}
\begin{proposition}
\label{NSequencesAreSufficient} If the rate vector $r$ has a
symmetric sum partition of size $N$, then $N$ orthogonal sequences
are sufficient to attain the minimum sum power $p_{tot}$ {=}
$\exp{\left\{2r_{tot}\right\}}-1$. Analogously, if the power
constraint vector $p$ has a symmetric sum partition of size $N$,
then $N$ orthogonal sequences are sufficient to attain the sum
capacity $r_{tot} {=} \frac{1}{2}\log\left(1+p_{tot}\right)$.
$\hfill \Box$
\end{proposition}

\vspace*{.1in}

\begin{proof}
We will prove the proposition for Problem II. A similar argument
holds for Problem I. Let $S_1,S_2,\cdots,S_N$ be the symmetric sum
partition of the users. An execution of the Algorithm
\ref{AlgorithmProblemII} that assigns sequences and powers to all
users in a subset $S_n$ before assigning to users in another subset
will result in an orthogonal allocation. This is because
\begin{eqnarray*}
  \exp{ \left\{ 2N \sum_{k \in S_n} r_k \right\} }
  =  \exp{ \left\{ 2r_{tot} \right\} }
  =  \lambda_{\max},
\end{eqnarray*}
which implies that when all users in $S_n$ are assigned, exactly one
dimension is completely filled to $\lambda_{\max}$. Users in one
subset will therefore be assigned the same sequence, and different
subsets are assigned orthogonal sequences. We will however prove the
proposition directly.

Let $S_1, S_2, \cdots, S_N$ be the symmetric sum partition of the
users. Assign sequences and powers as follows: if $k \in S_n$, then
\begin{eqnarray}
\label{seqassign} s_{k} & = & e_{n},
\\ \label{powerassign} p_{k} & = &  \frac{\mathfrak{I}_{k}(n)}{N} \left[ \exp \left\{ 2N r_{k} \right\} -1  \right],
\end{eqnarray}
where
\begin{eqnarray}
\label{I_kofndefn} \mathfrak{I}_{k}\left(n\right)
\stackrel{\Delta}{=} \exp \left\{ 2N \sum_{j:j \in S_{n}, j < k}
r_{j} \right\}
\end{eqnarray}
is the interference suffered by user $k$ due to presence of other
users in the same subset $S_n$ with a smaller index.

It is easy to see that
\begin{equation}
\label{achievablerate}
r_k {=} \frac{1}{2N}\log{\left(1+\frac{Np_k}{\mathfrak{I}_k\left(n\right)}\right)}\\
\end{equation}
is achievable via successive interference cancelation, where the
highest index user in this subset is decoded first. Users in other
subsets do not cause interference to users in this subset. Observe
that the total power assigned to users in any subset $S_n$ is given
by

\begin{eqnarray}
\label{subsetsumpower} \sum_{k \in S_{n}} p_{k} {=} \frac{1}{N}
\left(\exp \left\{ 2r_{tot} \right\} - 1 \right),
\end{eqnarray}
where (\ref{subsetsumpower}) follows by substitution of
(\ref{powerassign}) and (\ref{I_kofndefn}) in the left side of
(\ref{subsetsumpower}) and by observing that the resulting sum over
$S_n$ has only two terms that survive.

From (\ref{subsetsumpower}) we see that total power allocated to all
users is
\[
p_{tot} {=} \sum_{n{=}1}^{N}\sum_{k \in S_{n}} p_{k} = \exp \{ 2
r_{tot} \} -1,
\]
thus showing that $N$ orthogonal sequences are optimal.
\end{proof}

\vspace*{.1in}

Having derived a sufficient condition for optimality of $N$
orthogonal sequences, let us see how to manufacture this condition
from a given set of power constraints or rate requirements.

\vspace*{.1in}

\begin{proposition}
\label{splittingProposition}
Every strictly positive vector $x$
representing power constraints or rate requirements for $K$
non-oversized users can be cast into a vector $x'$ for $K'$ virtual
users, where $K \leq K' \leq K+N-1$, and $x'$ is such that it has a
symmetric sum partition of size $N$. Moreover $x'$ is obtained by
splitting $K'-K$ users into exactly two virtual users each. $\hfill
\Box$
\end{proposition}

\vspace*{.1in}

\begin{proof}
Consider the cumulative requirement
\[
  X_k {=} \sum_{i \leq k} x_i.
\]
$X_k$ is a strictly increasing function of $k$ and $X_K = x_{tot}$.
Hence there exist $N-1$ distinct users with indices $k_j$, $j =
1,2,\cdots,N-1$, such that
\begin{eqnarray}
X_{k_{j}-1} & = & \sum_{i = 1}^{ k_j - 1} x_i < \frac{jx_{tot}}{N},
\nonumber \\
\label{ratesum} X_{k_{j}} & = & \sum_{i{=}1}^{k_j} x_{i} \geq
\frac{jx_{tot}}{N}.
\end{eqnarray}
If strict inequality holds in (\ref{ratesum}), split user $k_j$'s
rate as

\begin{equation}
\label{splitrate}
   x_{k_{j}} =
   \underbrace{\left(\frac{j x_{tot}}{N}- X_{k_{j}-1}\right)}_{x_{k_j}^{'}}
   + \underbrace{\left(X_{k_{j}}-\frac{jx_{tot}}{N}
   \right)}_{x_{k_j}^{''}}.
\end{equation}
If equality holds in (\ref{ratesum}) leave the user as is. For users
that will be split, obtain $x^{'}$ from $x$ by replacing $x_{k_j}$,
the requirement for user $k_j$, by requirements ${x_{k_j}^{'}}$ and
${x_{k_j}^{''}}$ for two virtual users, where ${x_{k_j}^{'}}$ and
${x_{k_j}^{''}}$ are as in (\ref{splitrate}).

It follows that $x^{'}$ is a vector of size $K^{'}$, where $K \leq
K^{'} \leq K+N-1$, and $x^{'}$ has a symmetric sum partition of size
$N$.
\end{proof}

\vspace*{.1in}

Users whose rates or power constraints are split are assigned two
orthogonal sequences and will span a two dimensional subspace. The
design subsequently results in $N$ orthogonal sequences.

It is immediate that even for the case with oversized users, the
oversized users will be split into at most $N$ virtual users.
Non-oversized users will be split into at most two virtual users.
The resulting vector $r^{'}$ has at most $K+N-1$ elements and has a
symmetric sum partition of size $N$. Thus $N$ orthogonal sequences
are sufficient if rate or power splitting and multi-dimensional
signaling is allowed for some users. We make this precise in the
following Proposition. The proof is identical to the proof of
Proposition \ref{splittingProposition} and therefore omitted.

\vspace*{.1in}

\begin{proposition}
Every strictly positive vector $x$ representing power constraints or
rate requirements for $K$ users, regardless of the presence of
oversized users, can be cast into a vector $x^{'}$ of size $K^{'}$,
where $x'$ has a symmetric sum partition of size $N$, and $K \leq
K^{'} \leq K+N-1$. $\hfill \Box$
\end{proposition}

\section{Concluding Remarks}
\label{conclusions}

The work in this paper was primarily motivated by a desire to reduce
the amount of signaling necessary to communicate the sequences to
the geographically separated users. This signaling usually eats up
precious bandwidth in the downlink. This is especially a problem
when the channel changes, or when users enter or leave the system.
These significant events trigger a communication of a new set of
sequences and rates/powers in the downlink.

We showed that $2N - 1$ sequences are sufficient to attain maximum
sum capacity or minimum sum power. This results in some savings on
the downlink when the number of users are large. The base station
may broadcast the $2N -1$ sequences as common information and
indicate to each user only the index from the set and the rate or
power allocated. If there are oversized users, fewer sequences are
needed. The allocations are optimal and the algorithms have a simple
geometric interpretation. The algorithms take only $O(KN)$ steps,
have the same complexity as that of \cite[Algorithm
4]{200411TIT_TroDhiHea}, and the computations are numerically
stable. We also argued that there exist rate or power constraints
that necessitate $2N-1$ sequences.

We then saw that with a small penalty in the spreading factor for a
few users, $N$ orthogonal sequences are sufficient. The users and
the base station can then agree on a fixed orthogonal set of $N$
sequences. The base station only needs to signal the powers/rates
and the indices corresponding to the sequences allocated to a user.

The proposed algorithms identify a tight frame that meets the
supplied constraints. They may therefore find application in other
areas where tight frames arise.

The assumed multiple access channel model, however, has severe
limitations. The uplink wireless channel typically suffers from
multipath effects, fading, and asynchronism. Moreover, the users are
not active all the time. Yet, if the users can all be synchronized
via, for example, a Global Positioning System (GPS) receiver, our
results give some interesting design insights. For frequency-flat
slow fading channels where all the users with a tight delay
constraint have to be served simultaneously, multi-dimensional
signaling (more commonly referred to multi-code where the code is a
spreading sequence) can allow communication at sum capacity or
minimum power. Orthogonal sequences are sufficient and the signaling
in the downlink is significantly reduced. A successively canceling
decoder is necessary, but the complexity of this receiver is reduced
to a great extent because optimal decoding for the $K$ users
decouples into decoding for $N$ separate non-interfering groups.

Fairness of the allocation has not been considered in this paper.
However, with the rate splitting approach that separates virtual
users into groups, the decoding order of virtual users within a
group can be cycled to get a fairer allocation. The first user to be
decoded in a group treats all others in the group as interference
and suffers the most. Cycling ensures that this and other such
disadvantageous positions are shared in time by all users.

\appendices

\section{Proof of Proposition \ref{Interlacing}}

\label{proofOfInterlacingProposition}

\begin{proof}
Observe that the complex Hermitian matrix $A$ can be unitarily
diagonalized as follows
\[
  A = U \Lambda U^*,
\]
where $U = [u_1 ~ u_2 ~ \cdots ~ u_N]$ is the matrix of eigenvectors
of $A$ and $\Lambda = \mbox{diag} \{ \lambda_1, \cdots,
\lambda_N\}$, a diagonal matrix with real entries. Assuming that the
Proposition holds for real and diagonal matrices, we can find a real
vector $y$ such that $\sigma \left(\Lambda + y y^* \right) = \left(
\hat{\lambda}_1, \cdots, \hat{\lambda}_N \right)$. Clearly,
\[
  \sigma \left( U (\Lambda + y y^*) U^* \right) = \sigma \left( \Lambda + y y^*
  \right),
\]
and therefore the desired rank-1 perturbation of $A$ with the
prescribed eigenvalues is $A + \left( U y \right) \left( U y
\right)^*$, {\it i.e.}, $c = U y$. If $A$ is a real and symmetric
matrix, $U$ may be taken to be a real orthogonal matrix, and
therefore $c$ is a real vector.

It is therefore sufficient to prove the Proposition for diagonal
matrices with real entries. Let $A$ now be such a matrix. Let $y \in
{\mathbb{R}}^N$ be a nonzero vector. Let $y_N \neq 0$. The
characteristic polynomial $p_{\hat{A}}(t)$ of $\hat{A} = A + y y^T$
is computed as
\begin{eqnarray}
\lefteqn{ \nonumber \left| tI_N-\hat{A} \right| } \\
\nonumber & \stackrel{(a)}{=} & \left|
\begin{array}{cccc}
t-\lambda_1-y_1^2 & -y_1 y_2          & \cdots & -y_1 y_N \\
-y_2 y_1          & t-\lambda_2-y_2^2 & \cdots & -y_2 y_N \\
\vdots            & \vdots            & \ddots & \vdots   \\
-y_N y_1          & -y_N y_2          & \cdots & t-\lambda_N-y_N^2
\end{array}
\right| \\
\nonumber & \stackrel{(b)}{=} & \left|
\begin{array}{ccc}
t I_{N-1} - \Lambda_{N-1} & \vline & -y_N v \\

& \vline & \\

\cline{1-3}

& \vline & \\

-\frac{(t-\lambda_N)}{y_N} v^T & \vline & t - \lambda_N - y_N^2
\end{array}
\right| \\
\nonumber & \stackrel{(c)}{=} & \left|
\begin{array}{ccc}
I_{N-1} & \vline & {\bf 0} \\

& \vline & \\

\cline{1-3}

& \vline & \\

\frac{(t-\lambda_N)}{y_N} v^T(tI_{N-1} - \Lambda_{N-1})^{-1} &
\vline & 1
\end{array}
\right| \\
\nonumber && ~~~ \cdot \left|
\begin{array}{ccc}
t I_{N-1} - \Lambda_{N-1} & \vline & -y_N v \\

& \vline & \\

\cline{1-3}

& \vline & \\

-\frac{(t-\lambda_N)}{y_N} v^T & \vline & t - \lambda_N - y_N^2
\end{array}
\right| \\
\nonumber & & ~~~ \cdot \left|
\begin{array}{ccc}
I_{N-1} & \vline & y_N (t I_{N-1} - \Lambda_{N-1})^{-1} v \\

& \vline & \\

\cline{1-3}

& \vline & \\

{\bf 0}^T & \vline & 1
\end{array}
\right|\\
\nonumber & \stackrel{(d)}{=} & \left|
\begin{array}{ccc}
t I_{N-1} - \Lambda_{N-1} & \vline & {\bf 0} \\

& \vline & \\

\cline{1-3}

& \vline & \\

{\bf 0}^T & \vline & (t - \lambda_N) \left( 1 - \sum_{i=1}^N  y_i^2
\frac{1}{t - \lambda_i} \right)
\end{array}
\right| \\
& = & \label{CharacteristicPolynomial_Ahat} \prod_{i=1}^N (t -
\lambda_i) \left( 1 - \sum_{i=1}^N  y_i^2 \frac{1}{t - \lambda_i}
\right),
\end{eqnarray}
where $(b)$ follows by setting the columns of the matrix in $(a)$ to
be $h_i, i=1, \cdots, N$, then by replacing column $h_i$ by $h_i -
h_N y_i/y_N$, for $i = 1, \cdots, N-1$, and then by setting $v =
(y_1, \cdots, y_{N-1})^T \in {\mathbb{R}}^{N-1}$ and $\Lambda_{N-1}
= \mbox{diag}(\lambda_1, \cdots, \lambda_{N-1})$; $(c)$ follows by
multiplying the matrix in $(b)$ on either side by triangular
matrices that have determinant 1; $(d)$ follows by multiplying out
the matrices in $(c)$. Observe that the conclusion remains unchanged
if $y_N = 0$, but instead some $y_k \neq 0$ for some other $k$; such
a $k$ exists because $y$ is nonzero.

Define the $N$th degree polynomials
\begin{eqnarray}
\nonumber
f(t) & \stackrel{\Delta}{=} \prod_{i=1}^N (t - \hat{\lambda_i}), \\
\nonumber g(t) & \stackrel{\Delta}{=} \prod_{i=1}^N (t - \lambda_i).
\end{eqnarray}
Suppose now that the eigenvalues in $\lambda$ are all distinct. By
the Euclidean algorithm, we have
\[
f(t) = g(t) + r(t),
\]
where $r(t)$ is a polynomial of degree at most $N-1$. We also have
$f(\lambda_i) = r(\lambda_i)$ for $i=1,2, \cdots, N$ because
$g(\lambda_i) = 0$.

The polynomial $r(t)$ being known at $N$ different points can be
written explicitly by using the Lagrange interpolation formula
\[
r(t) = \sum_{i=1}^N f(\lambda_i)
\frac{g(t)}{g'(\lambda_i)(t-\lambda_i)}.
\]
Thus,
\begin{equation}
\label{CharacteristicPolynomial_Ahat_Alternate} \frac{f(t)}{g(t)} =
= 1 + \frac{r(t)}{g(t)} = 1 + \sum_{i=1}^N
\frac{f(\lambda_i)}{g'(\lambda_i)} \frac{1}{t-\lambda_i}.
\end{equation}
Comparing (\ref{CharacteristicPolynomial_Ahat}) and
(\ref{CharacteristicPolynomial_Ahat_Alternate}), it is clear that we
are done if we can set $y_i^2 \stackrel{\Delta}{=}
-f(\lambda_i)/g'(\lambda_i)$, for $i = 1, \cdots, N$. To do this,
however, we must show that $f(\lambda_i)/g'(\lambda_i) \leq 0$.
Interlacing property ensures that this is indeed the case. Indeed,
we can easily see that
\begin{eqnarray}
\nonumber
f(\lambda_i) & = & (-1)^{N-i+1} \prod_{j=1}^N |\lambda_i - \hat{\lambda}_j| \\
\nonumber g'(\lambda_i) & = & (-1)^{N-i} \prod_{j=1, j \neq i}^N
|\lambda_i - \lambda_j|,
\end{eqnarray}
and therefore $f(\lambda_i)$ and $g'(\lambda_i)$ are always of
opposite signs. Thus we may take
\begin{equation}
  \label{yVector}
  y_i = \sqrt{\frac{\prod_{j=1}^N |\lambda_i - \hat{\lambda}_j|}
  {\prod_{j=1, j \neq i}^N |\lambda_i - \lambda_j|}}.
\end{equation}

Next suppose that $\lambda$ has multiplicities. Observe that
$\lambda$ and $\hat{\lambda}$ interlace. Consequently, if the
spectrum $\lambda$ has multiplicity $k$ for a particular value
$\theta$, {\it i.e.}, $\lambda_{l+j} = \theta$ for some $l$ and for
$j=1, \cdots, k$, then the spectrum $\hat{\lambda}$ has multiplicity
$k-1$, or $k$, or $k+1$ for that $\theta$. This is because
$\hat{\lambda}_{l+j}, j=1, \cdots, k-1$, are pegged at $\theta$ due
to the interlacing property. So the multiplicity of $\theta$ is at
least $k-1$. It may further happen that $\hat{\lambda}_l = \theta$
or $\hat{\lambda}_{l+k} = \theta$, or both, leading to the other
possibilities.

We can therefore set
\[
\frac{f(t)}{g(t)} = \frac{\tilde{f}(t)}{\tilde{g}(t)},
\]
where the polynomials $\tilde{f}$ and $\tilde{g}$ have the same
degree $L$, with $1 \leq L \leq N$, and have no common factors.
Moreover, $\tilde{g}$ has no multiplicities and the zeros of
$\tilde{f}$ and $\tilde{g}$ interlace. The same arguments that lead
to (\ref{yVector}) hold with $y_{l_i}$ set as in (\ref{yVector}) for
those indices that survive and $y_j = 0$ when $j \neq l_i, i = 1,
\cdots, L$. The details are easy to fill and therefore omitted.
\end{proof}


\addcontentsline{toc}{section}{Acknowledgment}


\bibliography{InterlacingTheoremAndSequenceDesign.bib}

\begin{biographynophoto}{Rajesh Sundaresan}
(S'96-M'2000-SM'2006) received his B.Tech. degree in electronics and
communication from the Indian Institute of Technology, Madras, the
M.A. and Ph.D. degrees in electrical engineering from Princeton
University, NJ, in 1996 and 1999, respectively. From 1999 to 2005,
he worked at Qualcomm Inc., Campbell, CA, on the design of
communication algorithms for WCDMA and HSDPA modems. Since 2005 he
has been an Assistant Professor in the Electrical Communication
Engineering department at the Indian Institute of Science,
Bangalore. His interests are in the areas of wireless communication,
multiple access systems, information theory, and signal processing
for communication systems.
\end{biographynophoto}

\begin{biographynophoto}{Arun Padakandla}
received his B.E. degree in electronics and communication
engineering from the M. Visvesvaraya Institute of Technology,
Bangalore, in 2004. From 2004 to 2005 he worked at Motorola India
Electronics Ltd., Bangalore, on the implementation of modems for
iDEN cellular radio. Since 2005 he is pursuing a post graduation in
the Electrical Communication Engineering department at the Indian
Institute of Science, Bangalore. His interests are in multiuser
information theory, space time codes, and signal processing.
\end{biographynophoto}

\newpage

\newpage

\end{document}